\documentclass[draft]{agujournal2019}
\usepackage{url} 
\usepackage{lineno}
\usepackage[finalnew]{trackchanges} 
\usepackage{soul}

\draftfalse

\journalname{Space Weather}

\usepackage{aas_macros}
\soulregister\ref7
\soulregister\cite7

\begin{document}

%
%

\title{Assessing the Influence of Input Magnetic Maps on Global Modeling of the Solar Wind and CME-driven Shock in the 2013 April 11 Event}

%
%




\authors{M. Jin\affil{1, 2}, N. V. Nitta\affil{1},  C. M. S. Cohen\affil{3}}

\affiliation{1}{Lockheed Martin Solar and Astrophysics Lab (LMSAL), Palo Alto, CA 94304, USA}
\affiliation{2}{SETI Institute, Mountain View, CA 94043, USA}
\affiliation{3}{California Institute of Technology, Pasadena, CA 91125, USA}





\correspondingauthor{Meng Jin}{jinmeng@lmsal.com}




\begin{keypoints}
\item Due to the inhomogeneous background solar wind and dynamical evolution of the CME, the CME-driven shock parameters change significantly both spatially and temporally when the CME propagates in the heliosphere.
\item The input magnetic map has great impact on the shock connectivity and shock properties in the global MHD simulation, which is demonstrated in the 2013 April 11 event study. 
\item This study illustrates the model uncertainty due to the imperfect magnetic field observations, which should be considered when the model is used for research or space weather forecasting purposes.

\end{keypoints}

%
%

%
%


\begin{abstract}
In the past decade, significant efforts have been made in developing physics-based solar wind and coronal mass ejection (CME) models, which have been or are being transferred to national centers (e.g., SWPC, CCMC) to enable space weather predictive capability. However, the input data coverage for space weather forecasting is extremely limited. One major limitation is the solar magnetic field measurements, which are used to specify the inner boundary conditions of the global magnetohydrodynamic (MHD) models. In this study, using the Alfv\'{e}n wave solar model (AWSoM), we quantitatively assess the influence of the magnetic field map input (synoptic/diachronic vs. synchronic magnetic maps) on the global modeling of the solar wind and the CME-driven shock in the 2013 April 11 solar energetic particle (SEP) event. Our study shows that due to the inhomogeneous background solar wind and dynamical evolution of the CME, the CME-driven shock parameters change significantly both spatially and temporally as the CME propagates through the heliosphere. The input magnetic map has a great impact on the shock connectivity and shock properties in the global MHD simulation. Therefore this study illustrates the importance of taking into account the model uncertainty due to the imperfect magnetic field measurements when using the model to provide space weather predictions. 
\end{abstract}

\section*{Plain Language Summary}
As the origin of space weather, solar wind and coronal mass ejections (CMEs) play an important part in the space weather prediction. Similar as the terrestrial weather forecast, advanced models are developed driven by the available observations. However, the input data coverage for space weather forecasting is extremely limited. One major input data used to drive the solar wind and CME models is the solar surface magnetic field, for which the current telescopes can only observe less than half of the surface therefore assumptions are needed for the rest that leads to different types of magnetic maps. In this study, we quantitatively assess the influence of the two widely used magnetic maps on the global modeling of the solar wind the CME-driven shock in the 2013 April 11 event. Our result suggests that the input magnetic map has a great impact on the simulated background solar wind, CME-driven shock properties, as well as the spacecraft connectivity. Our study illustrates the importance of considering the model uncertainty due to the limited magnetic field coverage when using the model for research or space weather forecasting purposes.

%
%

\section{Introduction}
The broad topic of space weather represents the constantly changing physical conditions in the near-Earth environment, which are significantly influenced by the solar wind and coronal mass ejections (CMEs). Fast CMEs drive shocks in the corona and heliosphere \cite<e.g.,>[]{sime87, vour03} that are believed to be responsible for gradual solar energetic particle (SEP) events \cite<see reviews by>[]{reames99, desai16} primarily through the diffusive shock acceleration (DSA) mechanism \cite<e.g.,>[]{lee83, zank00}. Large SEP events can pose major hazards to technology and life in space. If the interplanetary CME is directed at Earth, its embedded magnetic field, especially in the presence of a strong southward-directed component $B_{z}$, can interact with Earth's magnetosphere and trigger non-recurrent geomagnetic storms \cite{gosling93}. Due to their critical importance to space weather prediction, significant efforts have been made in developing physics-based solar wind and CME models \cite<see reviews by e.g.,>[]{cranmer17, macneice18, gombosi18}. By specifying the radial component of the magnetic field at the inner (photospheric) boundary from observational magnetograms, these models can achieve a steady state solar wind and reproduce with some success the large-scale wind streams at 1 AU. Several simplified analytical flux rope models have also been developed to address the erupting magnetic structure of CMEs \cite<e.g.,>[]{titov99, gibson98, titov14, titov18}. For example, by initiating a Titov-D\'{e}moulin (TD) flux rope, \citeA{chip08} simulated the 2003 Halloween CME event and made a quantitative comparison between the synthetic coronagraph images and LASCO observations, in which the strong CME-driven shock was reproduced. \citeA{jin17a} simulated the 2011 March 7 CME event using the Gibson-Low (GL) flux rope with the Alfv\'{e}n wave solar model \cite<AWSoM;>[]{bart14} from the Sun to 1 AU and performed detailed comparisons with remote-sensing and in-situ observations. The results show that the simulation can reproduce many of the observed features near the Sun and in the heliosphere. A recent study by \citeA{torok18} simulated the famous 2000 July 14 ``Bastille Day" eruption using a modified TD (TDm) flux rope \cite{titov14} with the Magnetohydrodynamic Algorithm outside a Sphere model \cite<MAS;>[]{lionello13}. Starting from a stable magnetic flux rope and initiating the eruption by boundary flows, the simulation is able to reproduce the morphologies of the observed flare arcade, halo CME, and associated EUV wave and coronal dimmings. Although several differences are found between the simulated and observed flux rope at 1 AU, their model successfully captured core structure of the flux rope and the negative B$_{z}$ component, which led to a very strong geomagnetic storm. Note that there are other approaches for modeling the CME flux rope including the Non-linear Force Free Field extrapolation \cite<e.g.,>[]{wiegelmann04, wheatland06,  malanushenko14} and magnetofrictional models \cite<e.g.,>[]{ballegooijen00, yeates08, cheung12, jiang16}.

With increasing sophistication of data-driven models, we are taking steps toward achieving physics-based space weather forecast capability. However, the available input observations for solar wind and CME modeling are severely limited, thereby requiring (often significant) assumptions both for the ``missing data" and the physical conditions of the corona. The current data-driven Magnetohydrodynamics (MHD) solar wind models use the photospheric magnetic field observations as inputs. However, all the magnetic field observations (except those started only recently by \textit{Solar Orbiter}) are from the Sun-Earth line. From this perspective, we can adequately observe only less than one half of the solar surface, and need to make assumptions for the remaining areas, including the far-side and polar regions. There are two major types of magnetic maps that are widely used to drive the solar wind models. One is the \emph{synoptic or diachronic} magnetic maps that assembles 27-days of magnetic field observations into a single map. One issue with this approach is that the magnetic fields on the same diachronic magnetic map are observed at different times and thus the map contains data which are up to 27-days old. The other type is \emph{synchronic} magnetic maps that are based on surface flux transport models which simulate the evolution of the surface magnetic elements while assimilating new observations \cite<e.g.,>[]{schrijver03}. Additionally, a synchronic map may incorporate the newly emerged flux on the far-side of the Sun as inferred from helioseismology \cite{lindsey97, braun01, Hernandez07}. A preliminary study shows that by including the far-side flux in the Air Force Data Assimilative Photospheric Flux Transport (ADAPT) - Wang-Sheeley-Arge (WSA) model, the observed solar wind conditions could be better reproduced \cite{arge13}. However, doing so is not straightforward given the number of assumptions needed for the total fluxes and tilt angles of such emerging flux elements. Since most of the current solar wind models rely on either of these two types of magnetic maps, it is important to quantitatively evaluate the differences in ambient solar wind solutions as obtained using the diachronic orsynchronic magnetic map inputs. We note that some studies have examined the uncertainties of the solar wind solutions due to magnetic data from different observatories \cite{gressl14, riley14, jian15, hayashi16}. There are also recent studies that compare the solar wind solutions from the diachronic and synchronic magnetic inputs based on either PFSS model \cite<e.g.,>[]{wallace19, caplan21} or MHD models \cite<e.g.,>[]{linker17, lfw21}.

However, most of the previous studies have focused on comparing the ambient solar corona and wind solutions, and not transient structures, driven by the different magnetic inputs. It is widely known that the ambient solar wind can have significant influence on the evolution of the CME and the properties of the CME-driven shock wave, as has been studied both in simulations \cite<e.g.,>[]{riley99, riley03, ods04, jacobs05, hosteaux19} and in observations \cite<e.g.,>[]{gopalswamy00, temmer11}. Here, we propose to investigate the influence of different background solar wind solutions, due to the different magnetic inputs (i.e., diachronic vs. synchronic), on the CME properties in 3D. In particular, the parameters of the CME-driven shock (e.g., compression ratio, Mach number, shock speed, and shock angle $\theta_{B_n}$) are critical for understanding the particle acceleration in gradual SEP events. However, these shock parameters are difficult to determine directly from remote sensing observations \cite<e.g.,>[]{rouillard16, lario17, kwon18}. Furthermore, the field line connectivity plays an important role in understanding the in-situ SEP observations \cite<e.g.,>[]{lario17}, although it is also difficult to determine from current measurements. Therefore, to infer the link between the observer and the shock, most of the current SEP modeling efforts use Parker-spiral magnetic connectivity outside the source surface, which is often set at the heliocentric distance of 2.5~R$_{\odot}$. Due to the simplicity of this technique, such models do not always explain the observed characteristics of some SEP events \cite<e.g.,>[]{cairns20}. With the dynamic magnetic connectivity and shock parameters available from an MHD model, additional information relevant to the shock acceleration can be obtained through a physics-based approach.

In this study, we quantitatively assess the influence of the magnetic input, by modeling the CME on 2013 April 11 and its shock, which was associated with an SEP event \cite<e.g.,>[]{cohen14, lario14}. This article is organized as follows. In Section 2, we describe the models and methods used in this study, followed by results in Section 3 and discussion and conclusions in Section 4.

\section{Methodology}

\subsection{Data Description}
The CME was clearly associated with an M6.5 class flare starting at 06:55 UT in AR 11719 (N07E13, Carrington longitude $\sim$73$^{\circ}$). At the time, the Carrington longitude for Earth, STEREO A (STA), and STEREO B (STB) were $\sim$86$^{\circ}$, $\sim$219$^{\circ}$, and $\sim$304$^{\circ}$, respectively. We use both diachronic and synchronic magnetic maps based on SDO/HMI observations \cite{schou12} to specify the model's inner boundary condition of magnetic field. The synchronic magnetic map in use is maintained at the Lockheed Martin Solar and Astrophysics Laboratory and based on a flux transport model \cite{schrijver03}, which assimilates new observations within 60$^{\circ}$ from disk center. These magnetic maps are updated every six hours and can be downloaded directly from the \texttt{PFSS} package in \texttt{SSWIDL}. The latest documentation of the LM flux transport model can be found online (\protect\url{https://www.lmsal.com/forecast/surfflux-model-v2/}). The diachronic magnetic map is obtained from the Stanford HMI Carrington Rotation Synoptic Charts. At each Carrington longitude in the magnetic map, the data are averaged from 20 contributing magnetograms made within 2 hours of central meridian passage (i.e., $\pm$1.2$^{\circ}$ of the central meridian) with the outliers (values which depart from the median by 3$\sigma$) removed. See \citeA{liu17} for more details. We choose the HMI diachronic magnetic map with polar field correction \cite{sun18} accessible at JSOC (\url{http://jsoc.stanford.edu/}) under dataseries \texttt{hmi.synoptic\_mr\_polfil\_720s} (3600$\times$1440 resolution).  

\subsection{Solar Wind and CME Models}
The MHD solar wind model used in this study is the Alfv\'{e}n Wave Solar Model \cite<AWSoM;>[]{bart14}, which is a data-driven model with a domain starting from the upper chromosphere and extending to the corona and heliosphere. The AWSoM model has been implemented at NASA's Community Coordinated Modeling Center (CCMC). The inner boundary condition of the magnetic field can be specified by different magnetic maps as mentioned in \S 1.  The inner boundary conditions for electron and proton temperatures $T_{e}$ and $T_{i}$ and number density $n$ are set to be  $T_{e}=T_{i}=$ 50,000 K and $n =$ 2$\times$10$^{17}$ m$^{-3}$.  The fixed density and temperature at the inner boundary do not otherwise have an evident influence on the global solar corona and wind solution \cite{lio09}. The detailed model validation on coronal density has been conducted by comparing numerical results with spectral line (e.g., EIS and SUMMER) observations \cite{oran13}, EUV line intensities \cite{jin17a}, and more recently through the density derived from the Differential Emission Measure Tomography \cite{sachdeva19}. The Parker solution \cite{parker58} is used to specify the initial conditions for the solar wind plasma, while the initial magnetic field is based on the Potential Field Source Surface (PFSS) model with the Finite Difference Iterative Potential Solver \cite<FDIPS;>[]{toth11}. In this study, the source surface is set at 2.5 R$_\odot$. The global solar wind solution is obtained by coupling the solar corona (SC; from 1 to 24 R$_{\odot}$) and inner heliosphere (IH; from 18 to 250 R$_{\odot}$) components within the Space Weather Modeling Framework \protect\cite<SWMF;>[]{toth12}.

Alfv\'{e}n waves are prescribed as outgoing Alfv\'{e}n wave energy density that scales with the surface magnetic field. The solar wind is heated by a phenomenological description of Alfv\'{e}n wave dissipation and accelerated by thermal and Alfv\'{e}n wave pressure. Electron heat conduction (both collisional and collisionless) and radiative cooling are also included in the model, which are important for creating the solar transition region self-consistently. In addition, the model electron and proton temperatures are treated separately for producing physically correct solar wind and CME structures (e.g., CME-driven shocks), in which the electrons and protons are assumed to have the same bulk velocity but heat conduction is applied only to electrons due to their much higher thermal velocity \cite{chip12, jin13}. By introducing the phenomenological description of Alfv\'{e}n wave dissipation as well as the wave reflection and heat partitioning between the electrons and protons based on the results of linear wave theory and stochastic heating \cite{chandran11}, the AWSoM model has demonstrated the capability to reproduce the solar corona environment with three free parameters that determine the Poynting flux ($S_A / B$), the wave dissipation length ($L_\perp\sqrt{B}$), and the stochastic heating parameter ($h_S$) \cite{bart14}. 

To initiate a CME eruption, we use the Gibson-Low (GL) flux rope model \cite{gibson98} which has been successfully used in numerous modeling studies of CMEs \cite<e.g.,>[]{chip04a, chip04b, lugaz05a, lugaz05b, schmidt10, chip14}. Analytical profiles of the GL flux rope are obtained by finding a solution to the magnetohydrostatic equation $(\nabla\times{\bf B})\times{\bf B}-\nabla p-\rho {\bf g}=0$ with the solenoidal condition $\nabla\cdot{\bf B}=0$. To get the solution, a mathematical stretching transformation $r\rightarrow r-a$ is applied to an axisymmetric, spherical ball of twisted flux $\bf b$ with $r_0$ diameter centered at $r=r_1$ relative to the heliospheric coordinate system. The field of $\bf b$ can be expressed by a scalar function $A$ and a free parameter $a_1$ that determines the magnetic field strength \cite{lites95}. The flux rope acquires a tear-drop shape of twisted magnetic flux after the transformation. Also, Lorentz forces are introduced that lead to a density-depleted cavity in the upper portion and a dense core at the lower portion of the flux rope. This flux rope structure mimics the 3-part density structure of the CME seen in observations \cite{llling85}. The GL flux rope profiles are then superposed onto the steady-state solar wind solution: i.e. $\rho=\rho_{0}+\rho_{\rm GL}$, $p=p_{0}+p_{\rm GL}$, ${\bf B = B_{0}+B_{\rm GL}}$. The combined background-flux rope system is in a state of force imbalance, and thus erupts immediately when the simulation is advanced forward in time. The GL flux rope is mainly controlled by five parameters: the stretching parameter, $a$, determines the flux rope shape; the distance of torus center from the center of the Sun, $r_{1}$, determines the initial position of the axisymmetric flux rope before it is stretched; the radius of the flux rope torus, $r_{0}$, determines the flux rope size; the flux rope field strength parameter, $a_1$, determines the magnetic field strength of the flux rope; and a helicity parameter to determine the positive (dextral) /negative (sinistral) helicity of the flux rope \cite{jin17b, borovikov17}. \citeA{jin17b} developed a new method, Eruptive Event Generator Gibson-Low (EEGGL), to calculate GL flux rope parameters through a handful of observational quantities (i.e., magnetic field of the CME source region and observed CME speeds from white-light coronagraphs) so that the modeled CMEs can propagate with the desired CME speeds near the Sun.

\subsection{Summary of Approach}
To quantitatively assess the influence of magnetic input on the global modeling, we choose two different magnetic maps: the Lockheed Martin (LM) synchronic magnetic map for 2013 April 11 06:04:00 UT (referred to as input to Case I) and the diachronic Carrington magnetic map of CR2135 (referred to as input to Case II). Both types of magnetic maps have been widely used in the solar and heliospheric physics community for space weather nowcast/forecast purposes. The original magnetic maps are first resized to 360$\times$180 resolution that matches simulation grid while preserving the flux. For the LM synchronic map, this is done directly through the \texttt{PFSS} package in \texttt{SSWIDL}. For the HMI diachronic map, we resize the data from the original map in 3600$\times$1440 resolution. In addition, both magnetic maps have flux imbalance (i.e., zero-point error), which is calculated to be -6.9$\times$10$^{21}$ Mx ($\sim$-2\% of the total unsigned flux) and 5.4$\times$10$^{21}$ Mx  ($\sim$1.3\% of the total unsigned flux) for the LM synchronic map and HMI diachronic map respectively. This zero-point error is corrected by removing the average field (i.e. monopole) from the original magnetic maps \cite{toth11}. Other than this correction, we do not apply any scaling factor to the input magnetic maps in this study.

\begin{figure}
\centering
\includegraphics[width=1.0\textwidth]{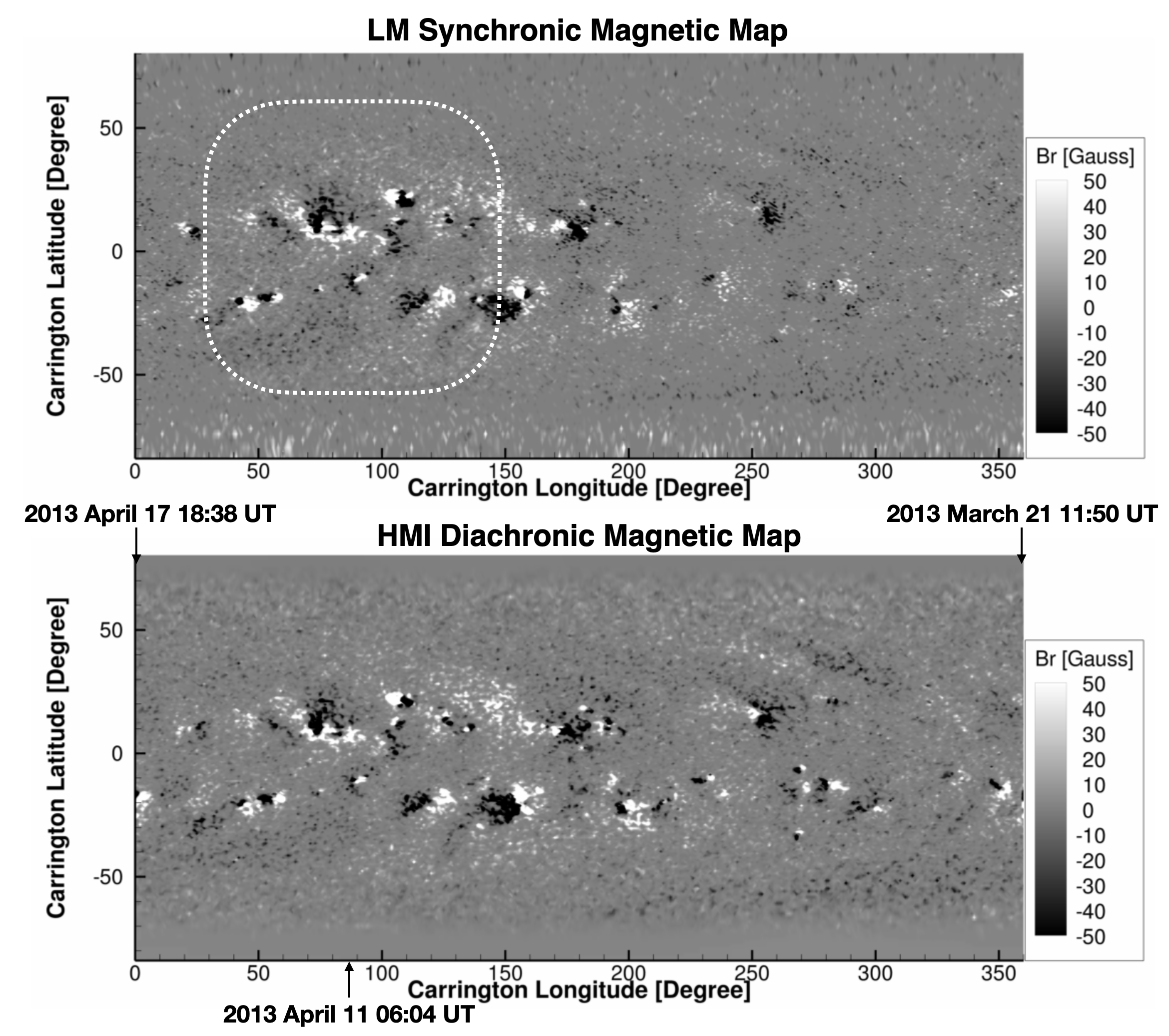}
\caption{The Lockheed Martin synchronic magnetic map on 2013 April 11 (upper panel) input for Case I and the HMI diachronic magnetic map of Carrington Rotation 2135 (lower panel) input for Case II. The white dotted box shows the area with assimilated HMI observations. The dates on the diachronic map show the  start/end times of the Carrington Rotation 2135 and the location where 2013 April 11 data is included.}
\label{magnetograms}
\end{figure}

Figure~\ref{magnetograms} shows the two magnetic maps used in this study. The LM synchronic magnetic map contains magnetic field observations only $\pm$60$^{\circ}$ from the disk center on 2013 April 11 06:04:00 UT (marked with white dotted box in Figure~\ref{magnetograms}) while the rest of the magnetic map is based on the flux transport model. Due to the different methods used to produce these magnetic maps mentioned in \S 2.1, one can see evident differences between the two magnetic maps for areas with longitude $>$200$^{\circ}$. However, the magnetic fields around the source region ($\sim$73$^{\circ}$) are similar between the two maps. To quantitatively compare the two magnetic maps, we further calculate the unsigned flux for both the whole map and the assimilating window, and the mean polar field strength. The results are summarized in Table~\ref{summarytable}. First, we need to note that for newly assimilated observation (marked as white dashed window in Figure~\ref{magnetograms}) in the LM synchronic map, the HMI flux is multiplied by a factor of 1.4 in order to match the previous MDI observations \cite{liu12}. Even with that enhanced magnetic flux, the total unsigned flux in the synchronic map (3.5$\times$10$^{23}$ Mx) is still $\sim$17\% less than that in the diachronic map (4.1$\times$10$^{23}$ Mx), which is mainly due to more and stronger magnetic structures involved in the diachronic map outside the assimilating area of the synchronic map where the flux diffuses in the flux transport model. However, we find that the total unsigned flux within the assimilating window is similar between the two maps (1.3$\times$10$^{23}$ Mx for synchronic map vs. 1.2$\times$10$^{23}$ Mx for diachronic map). Considering the factor of 1.4 applied to the field in the assimilating window of synchronic map, this also means there must be considerable field evolution around 2013 April 11 (e.g., newly emerged fluxes after 2013 April 11). Note that based on the start/end times of Carrington Rotation 2135, the assimilating window area in the synchronic map corresponds to $\sim$9 days of observation (2013 April 6 to 2013 April 15) in the diachronic map. On the other hand, similar unsigned flux also means that about the same amount of Poynting flux is initiated in the AWSoM model for heating the corona. Another noticeable difference between the two maps is the polar field (Carrington latitude $>$ 60$^\circ$). For LM synchronic map, the polar field is simulated from the flux transport model over many solar rotations. For the HMI diachronic map used here, the polar field is extrapolated from previous observations when part of the polar region could be seen due to the Sun's tilt angle. Therefore, we can see many small-scale structures in the polar region of the synchronic map while the polar field in the diachronic map is smoothed due to the extrapolation. Nevertheless, we found that the mean magnetic fields in the south polar region are very similar (2.7 Gauss for synchronic map vs. 2.6 Gauss for diachronic map). For the north polar region, the synchronic map has a lower mean field of -1.5 Gauss comparing with the -2.2 Gauss field in the diachronic map.

We run two steady-state simulations with the inner boundary condition of the magnetic field specified by the two different magnetic maps. After reaching the steady-state, we initiate a CME eruption by inserting a Gibson-Low flux rope into both steady-state solutions and integrate the model equations forward in time. The GL flux rope parameters are identical in the two simulation cases. We run the two cases for a duration of one hour in total simulated time, trace the shock location in 3D and calculate the shock parameters in order to compare the two simulations. 

The 3D shock front is determined by computing the derivative of entropy along the radial rays originating from the center of the Sun. The entropy is evaluated as $s=ln(T_p/\rho^{\gamma-1})$  where $T_{p}$ is the proton temperature, $\rho$ is the plasma density, and $\gamma$ is the polytropic index {($\gamma$=5/3)}. Once the shock front is determined, the shock normal ${\bf n}$ is calculated by using the magnetic coplanarity condition ${\bf (B_{2}-B_{1})}\cdot {\bf n}=0$ \cite{lepping71, abraham72}:
\begin{equation}
{\bf n}=\pm\frac{({\bf B_{2}}\times{\bf B_{1}})\times({\bf B_{1}}-{\bf B_{1}})}{|({\bf B_{2}}\times{\bf B_{1}})\times({\bf B_{2}}-{\bf B_{1}})|}
\label{eq_shock}
\end{equation}
where 1 and 2 represent the shock downstream and upstream conditions, respectively. The shock speed is determined by the conservation of mass across the shock: $v_{s} = \frac{\rho_{2}u_{2n}-\rho_{1}u_{1n}}{\rho_{2}-\rho_{1}}$. The shock Alfv\'{e}n Mach number is defined as $M_{A} = v_{s}/v_{A}$, where $v_{A}$ is the local Alfv\'{e}n speed. The shock angle $\theta_{B_n}$ is obtained by measuring the angle between the upstream magnetic field and the shock normal. To get the shock connectivity to different spacecraft locations, we extract field lines from the outer boundary of SC at 24 R$_\odot$ to the surface of the Sun. Since in this study, we did not extend the domain to include the IH component for the CME simulation, the connectivity from the spacecraft locations to the outer boundary of SC is determined by the steady-state solution that did include the heliospheric domain. Considering the relative short simulation time, this simplification has minimal influence on our results. To get the shock evolution profiles, for each time step (in 1 minute temporal resolution), we extract the shock parameters on the shock surface closest to the spacecraft-connecting field lines. 

\section{Results}
\subsection{Comparison of Steady-state Solutions}
In order to compare the two steady-state solar wind solutions constructed using the two magnetic maps, we calculate the locations of the positive (marked in green) and negative (marked in purple) open flux from both MHD and PFSS solutions as shown in Figure~\ref{steady-state}. The open flux is identified by tracing the field lines from the outer boundary at 24 R$_{\odot}$ in the simulations in a uniform latitude/longitude grid with one degree resolution back to the surface of the Sun. To quantitatively compare the results, we also calculate the open field area for both the positive and negative polarities and the results are summarized in Table~\ref{summarytable}. For both of the magnetic maps, the MHD and PFSS solutions are quite similar with the total open area slightly higher in the PFSS solution. The similarity between the MHD and PFSS solutions also suggests that the inherent properties of the different magnetic maps play a major factor in generating the different topological features instead of other model-related parameters. However, we need to note that the statistics in this particular case may not be generalized for all the MHD/PFSS models. Other adjustable parameters of both MHD and PFSS models could influence the open field area. For example, a lower source surface radius in the PFSS model could lead to larger open field area \cite{caplan21}, while the coronal heating (i.e., Alfv\'{e}n wave heating) related parameters in the MHD model could also influence the field opening \cite{linker17}. Comparing between the synchronic and diachronic map, both MHD and PFSS solutions show similar total open field area. However, there are also noticeable differences: For example, the positive flux area is larger in Case II than in Case I, while the negative flux area is larger in Case I.

\begin{figure}
\centering
\includegraphics[width=1.0\textwidth]{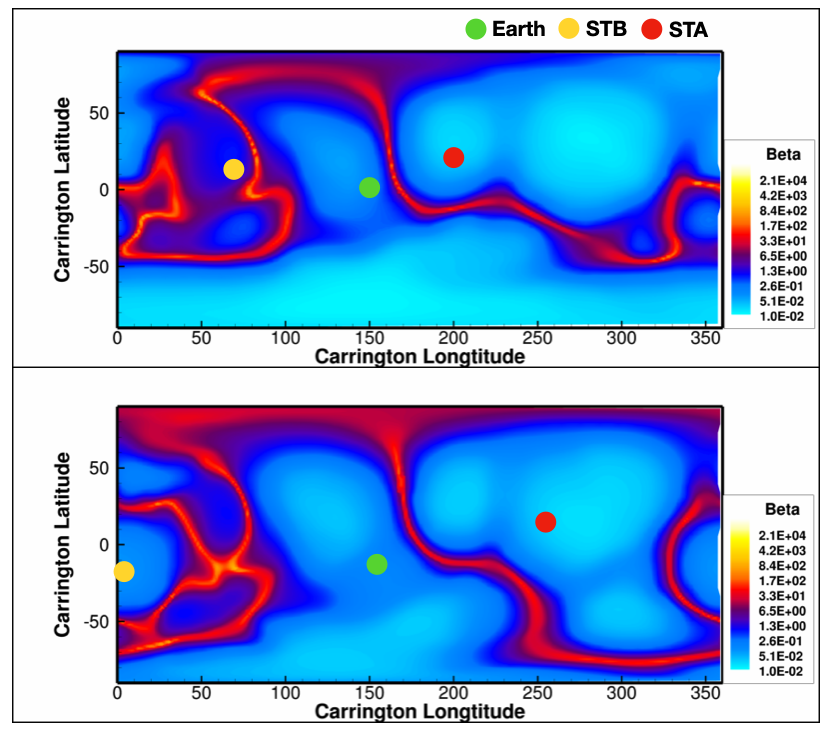}
\caption{Plasma-$\beta$ at 2.5 R$_{\odot}$ for MHD solutions using LM synchronic magnetic map (upper panel) and diachronic magnetic map (bottom panel).  The footpoints of Earth, STA, and STB are shown as green, red, and yellow dots.} 
\label{hcs}
\end{figure}

We extract the field lines connecting to Earth, STEREO A (STA), and STEREO B (STB) and mark the footpoints on the magnetic maps (indicated with colored circles). It is also apparent in Figure~\ref{steady-state} that the connectivity to the three 1 AU locations are quite different between the two solutions. For Earth, although both calculated footpoints are around the same longitude, they are $\sim$20$^\circ$ different in latitude. For STA, the footpoints are at approximately the same latitude in northern hemisphere, but they differ by $\sim$50$^\circ$ in longitude. The largest difference is found in the STB footpoints, which vary by $\sim$70$^\circ$ in longitude and $\sim$40$^\circ$ in latitude. Moreover, the polarities are different for the two STB footpoints. In Figure~\ref{hcs}, the plasma-$\beta$ at 2.5 R$_{\odot}$ from the MHD solutions are shown for the two cases; the location of the heliospheric current sheet (HCS) is indicated by the high plasma-$\beta$ values. Although there are a number of differences between the two cases, of note is that STB’s footpoint shifts from one side of the HCS to the other (this is not the case for STA or Earth). This connectivity difference significantly influences the shock profile evolution after the CME eruption observed at each location, as is discussed in the next section.

\subsection{Comparison of CME-driven Shocks}
\begin{figure}
\centering
\includegraphics[width=1.0\textwidth]{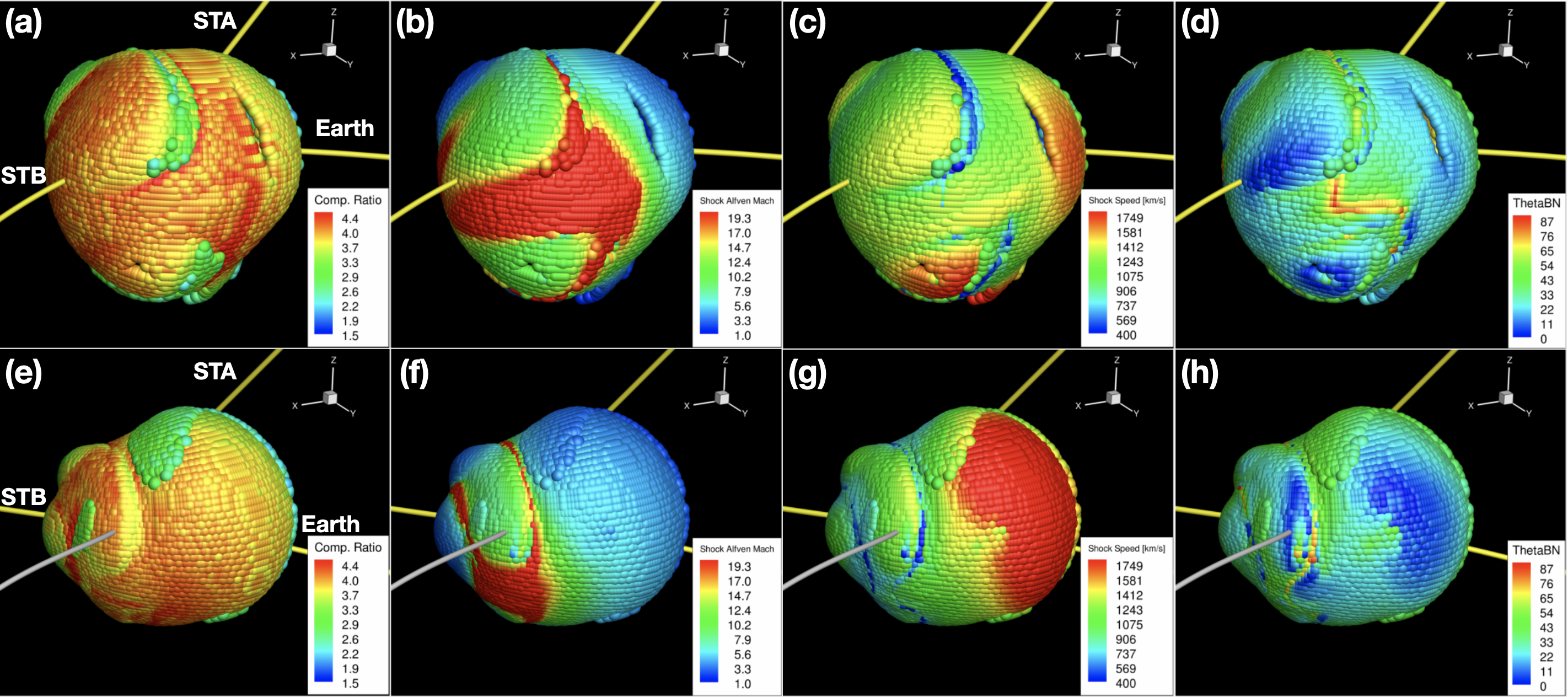}
\caption{The shock parameters along the shock surface at $t$ = 30 minutes in the two CME simulations. (a)-(d): CME-driven shock parameters derived from Case I using LM synchronic magnetic map; (e)-(h): CME-driven shock parameters derived from Case II using HMI diachronic magnetic map. The yellow field lines represent the connectivity to Earth, STA, and STB. The white field lines in (e)-(h) represent a field line rooted in the same region as STB connecting field line in Case I.}
\label{3d_shock}
\end{figure}

With the same CME flux rope running through the two different ambient coronal and solar wind solutions, we obtain two CME simulation cases. The Cartesian coordinate system used in the simulation is the heliographic rotating coordinates (i.e., Carrington coordinates) with the X-axis pointing to the Carrington longitude 0$^{\circ}$ and Y-axis pointing to the Carrington longitude 90$^{\circ}$. In Figure~\ref{cme_bg}, we show radial velocity field, total magnetic field strength, and plasma density on the $X$ = 0 and $Z$ = 0 planes at $t$ = 30 minutes for the two CME simulation cases, from which we can also see the background solar wind conditions based on the two magnetic maps.  In general, there are similar patterns we can identify between the two cases. However, there are also evident differences in the two solar wind conditions that lead to different shock parameters as discussed below.
In Figure~\ref{3d_shock}, we show the 3D shock parameters (compression ratio, shock Alfv\'{e}n Mach number, shock speed, and shock angle $\theta_{B_n}$) extracted from the two cases at $t=$ 30 minutes. The yellow field lines represent the connectivity to Earth, STA, and STB. Due to the inhomogeneous background solar wind, the CME-driven shocks in both cases are highly structured in shape and all the shock parameters vary significantly across the shock surface. By comparing the two cases with different magnetic field inputs, we can see several major differences: 1) The morphology of the shock surface, in that the latitudinal expansion is larger in Case I than in Case II; 2) The spatial distribution of Mach number along the shock surface is evidently different between the two cases. We find that the high Mach numbers in both cases are due to the smaller local Alfv\'{e}n speeds in the shock upstream as shown in Figure~\ref{upstream}b and d (area indicated by white arrows). These smaller Alfv\'{e}n speeds are related to both low magnetic field strength (Figure~\ref{cme_bg}b and e) and enhanced plasma density (Figure~\ref{cme_bg}c and f); 3) The spatial distribution of shock speed differs. The larger shock speed found at the leading front of the shock surface in Case II is related to the faster background solar wind around that region (marked by white arrows in Figure~\ref{cme_bg}a and d). See also the higher solar wind speed in the upstream of the shock in Case II (Figure~\ref{upstream}a and c). Finally, the footpoint differences seen in Figure~\ref{steady-state} result in the Earth, STA, and STB being connected to different parts of the shock in the two cases. 

\begin{figure}
\centering
\includegraphics[width=1.0\textwidth]{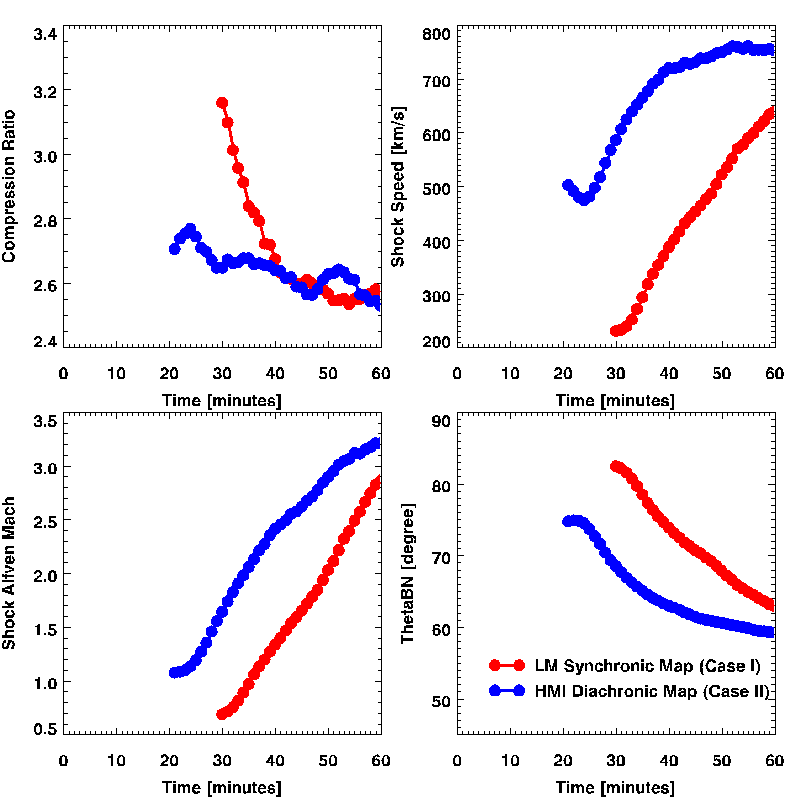}
\caption{Evolution of shock parameters connected to the Earth location in the two cases with different magnetic field inputs.}
\label{earth_profile}
\end{figure}

\begin{figure}
\centering
\includegraphics[width=1.0\textwidth]{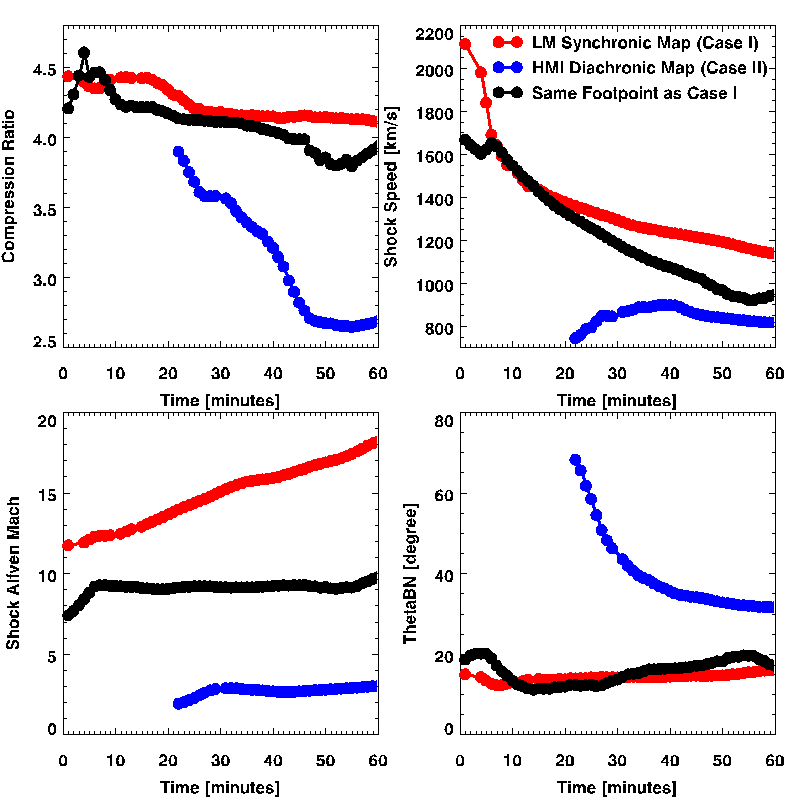}
\caption{Evolution of shock parameters connected to the STB location in the two cases with different magnetic field inputs. The black curves represent the shock evolution with a field line extracted from Case II that rooted from the same area as the STB connecting field lines in Case I.}
\label{stb_profile}
\end{figure}

To quantitatively evaluate the differences of the CME-driven shock parameters in the two cases, we derive the evolution of shock parameters connected to the Earth and STB locations in the two cases and show the results in Figure~\ref{earth_profile} and \ref{stb_profile}. Note that no shock connection is developed for STA in either cases and therefore, the STA profiles are not shown. The temporal resolution of the shock profiles is 1 minute. We found that the shock compression ratio calculated for STB location in Case I is slightly larger than 4 (strong shock limit), which is due to the nonideal process (e.g., heat conduction) involved in the MHD model. Also, we need to note that due to the unstable flux rope insertion used in this study, the shock parameters derived at the beginning could be unrealistically stronger, especially right in front of the flux rope driver. However, as the flux rope starts interacting with the global corona, the resulting CME tends to acquire a speed that agrees with the observations reasonably well as found in our previous studies \cite{jin17a, jin18}.

As we mentioned before, the footpoints connected to Earth in the two cases are around the same heliospheric longitude; this results in similar trends in the evolution of the shock parameters for the two cases. Also, the shock has a perpendicular nature in both cases throughout the first hour of the simulation. However, comparing the absolute values of the shock parameters, they are quite different between the two cases. In addition, the shock connection time is different in the two cases; in Case I, the connection is developed $\sim$10 minutes later than in Case II with the connection to Earth established only $\sim$20 minutes after the eruption onset. 

In contrast, due to the large difference in the locations of the STB footpoints, the CME-driven shock properties connecting to STB in the two cases are significantly different as shown in Figure~\ref{stb_profile}. The shock connecting to STB in Case I is a much stronger shock than that in Case II with a much higher compression ratio, shock Alfv\'{e}n Mach number, and shock speed. Also, the shock in Case I has a parallel nature (i.e., shock angle $<$ 20$^\circ$) while the shock in Case II is more oblique and has a perpendicular nature in the initial $\sim$10 minutes after first contact. Furthermore, the connection in Case I is established $\sim$20~minutes earlier than in Case II, which might have an effect on the properties of the resulting SEP event. For example, based on the DSA theory \cite{drury83}, the instantaneous particle spectral index depends only on the shock compression ratio and the higher shock compression ratio in Case I will lead to a harder SEP spectra. The higher shock compression ratio and quasi-parallel nature found in the Case I also suggests the shock is a more efficient accelerator than in the Case II \cite{ding20}. However, due to the complex physical processes of the particle acceleration and transport involved, to quantitatively link the shock properties near the Sun to the SEP spectra at 1 AU requires advanced coupling between the MHD model and a particle acceleration/transport model \cite{young21, li21}, which is beyond the scope of this work and will be examined in a future study.

As shown in Figure~\ref{3d_shock}, one major reason for the completely different shock profiles connecting to STB is the field line connectivity in the two cases, which results in STB being connected to different parts of the shock. In Case I, STB connects more closely to the front of the shock, while in Case II the connection is closer to the shock flank. To make a more comprehensive comparison, we also extract the field line with a footpoint rooted in the same region as the field line connecting to STB in Case I (shown as a white field line in Figure~\ref{3d_shock}e-h). The shock evolution profile from this extra field line is overlaid in Figure~\ref{stb_profile} (marked in black). We can see that when comparing shock profiles around the similar longitude/latitude location in the two cases, the shock parameters and their evolution are more similar mainly due to the same flux rope driver initiated in the two simulations. However, we want to emphasize that this similarity does not otherwise reconcile the issue we raised in this study as the different spacecraft connectivity in the two cases is largely attributable to different input magnetic maps. Furthermore, even at a similar location on the shock surface, certain shock parameters (e.g., Shock Alfv\'{e}n Mach number) still show clear differences between the two cases, which is caused by the different background solar wind solutions.

\section{Discussion \& Conclusions}
In the previous section, we have shown that the CME-driven shock parameters and connectivity can be significantly different when different types of magnetic maps are used to drive global MHD models. Here, we briefly discuss how these differences could influence our interpretation of observations in the 2013 April 11 SEP event. 

\begin{figure}
\centering
\includegraphics[width=1.0\textwidth]{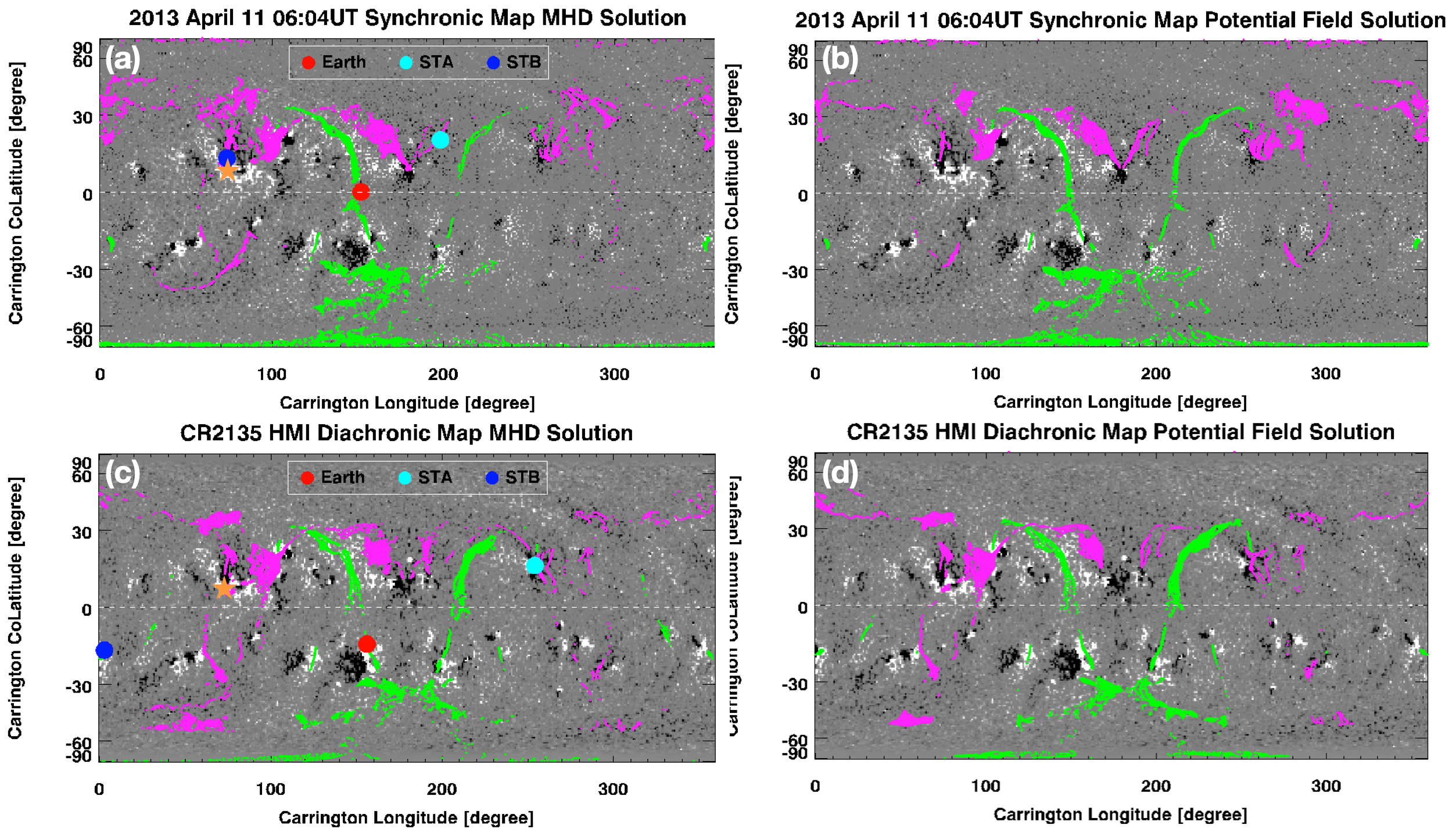}
\caption{The positive (green) and negative (purple) open flux calculated from the MHD and PFSS solutions using (Case I) LM synchronic magnetic map (a-b) and (Case II) HMI CR2135 diachronic magnetic map (c-d). The footpoints of field lines connecting to Earth, STA, and STB locations in the MHD solutions are also shown as red, cyan, and blue dots. The star marks the flare location.}
\label{steady-state}
\end{figure}

As shown in Figure~\ref{steady-state}, one major difference found in field line connectivity between the two cases is the footpoint connecting to STB. In Case I, we can see that the STB-connecting field line can be traced back to the flare site, while in Case II, the footpoint is far from the flare site. The SEP observations of this event show that the Fe/O ratio is higher at STB than at Earth \cite{cohen14}. Based on Case I, one possible explanation for the high Fe/O ratio could be that there is a direct contribution from the flare-accelerated, Fe-rich material \cite{cane03, cane06}. However, Case II does not support this interpretation as the STB and Earth footpoints have about the same longitudinal separation from the source region and therefore presumably more likely to measure similar Fe/O ratios.

Another feature that can be compared with observations is the shock-connection time. We found that in Case I, STB develops connection to the shock $\sim$30 minutes earlier than Earth, while in Case II, STB and Earth develop connections to the shock around the same time ($\sim$20 minutes after the eruption). The shock-connection time can be related to the particle release time calculated from the SEP in-situ observations. For this event in particular, \citeA{lario14} found that the estimated proton release time at STB from velocity dispersion analysis is 07:10 UT$\pm$4 minutes and 07:58 UT$\pm$9 minutes at Earth, which suggests that STB developed a connection to the shock earlier than the Earth did by 48$\pm$13 minutes, consistent with the $\sim$30 minutes found in Case I.

However, there are also features the two cases agree on. For example, in both cases, the shock connecting to STB is stronger than the shock connecting to Earth. This feature is consistent with the SEP observations that the energy spectra of He, O, and Fe are harder at STB than at Earth \cite{cohen14}. Also, the shock geometry connecting to the Earth location has a quasi-perpendicular nature (i.e., shock angle $>$45$^{\circ}$) in both cases, although the shock is more oblique in Case I. For the shock connecting to STB, although the shock starts with a quasi-perpendicular nature in Case II, the shock angle quickly decreases with time. After $\sim$40 minutes, the shock in both cases has a quasi-parallel nature with Case II more oblique.

We need to note that in this study, we are mainly focused on the influence of input magnetic maps on the MHD model output. But there are other parameters that could influence the modeling result. For example, the coronal heating parameters used in the MHD model (e.g., input wave energy density and length scale of energy dissipation) can have great influence on the final field topology \cite{linker17}. The enhanced coronal heating or shorter dissipation length could lead to more field opening in the MHD simulation. For the PFSS model, when using a lower source surface location, it also leads to more open field in the solution \cite{caplan21}. Moreover, the solar corona is a continuously changing environment, which may not be correctly represented by either PFSS model or a relaxed steady-state MHD solution. In contrast, the global magneto-friction model could provide an approach to account for such time-dependent factors \cite{yeates10, cheung12, fisher15}. Last but not least, the CME flux rope models used for initiating the eruptions could also play an important role for different CME-driven shock properties. Therefore, more work is needed for a comprehensive assessment on these factors and their relative importance on the model output.

\begin{figure}
\centering
\includegraphics[width=1.0\textwidth]{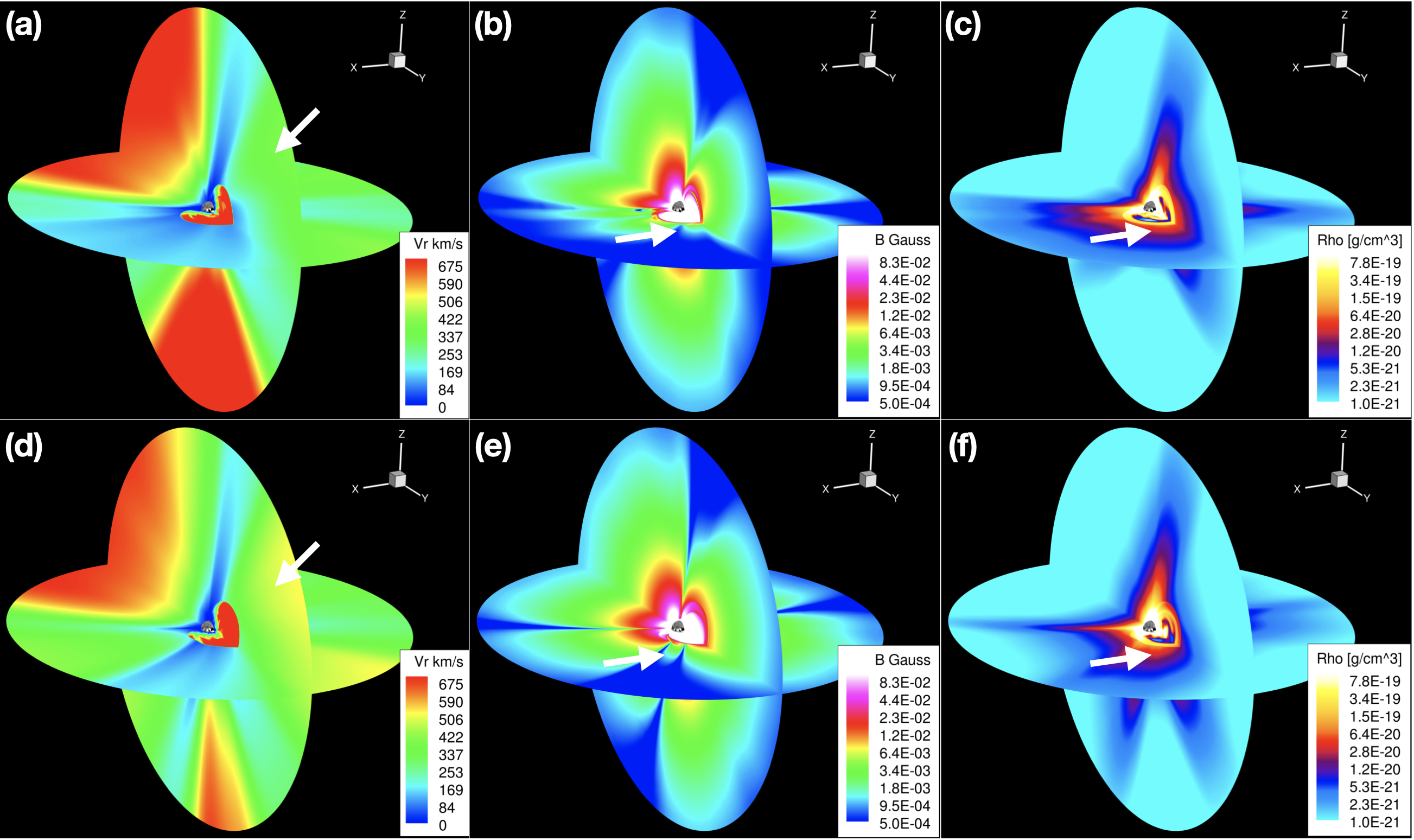}
\caption{(a) Radial velocity, (b) magnetic field strength, and (c) plasma density on the $X$ = 0 and $Z$ = 0 planes at $t$ = 30 minutes in the Case I CME simulation; (d)-(f): Same as (a)-(c) for the Case II CME simulation. The outer boundary is at 24 R$_{\odot}$. The white arrows in the figure point to the areas discussed in \S 3.}
\label{cme_bg}
\end{figure}

\begin{figure}
\centering
\includegraphics[width=1.0\textwidth]{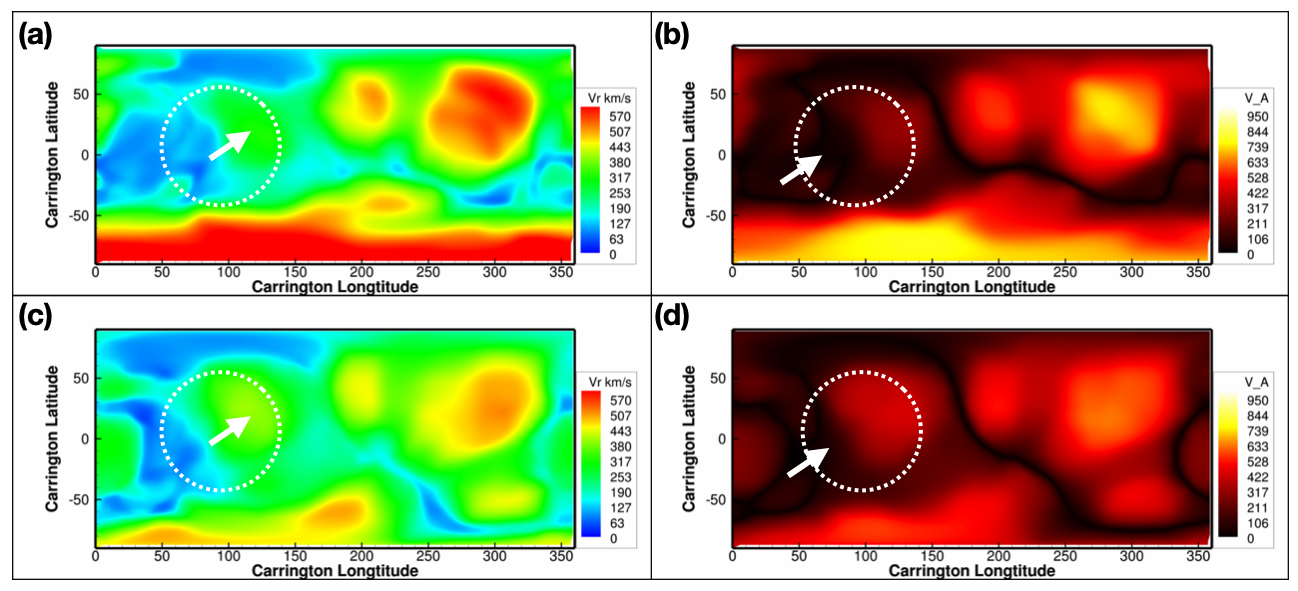}
\caption{(a) Ambient solar wind radial velocity and (b) Alfv\'{e}n speed at 7.5 R$_{\odot}$ in the Case I solution; (c)-(d): Same as (a)-(b) for Case II. The white dashed circles show approximate CME area and the white arrows point to the areas discussed in \S 3.}
\label{upstream}
\end{figure}

In this study, using the CME on 2013 April 11 associated with a SEP event as an example, we demonstrated how the choice of magnetic inputs can result in significantly different simulated background solar wind and, therefore, the CME-driven shock parameters and spacecraft connectivity in a global MHD model. There are multiple differences in the details of the synchronic and diachronic maps that may contribute to the differences in the results of the two cases:  the characteristics of the active regions (especially on the far-side), the strength of the polar field, and the flux imbalance between the polar regions. For the case in this study, we found that the flux imbalance between the two polar regions is much larger for the LM synchronic map (ratio of 1.8) than for the diachronic map (ratio of 1.2). This flux imbalance could influence the global field topology, position of the HCS, and magnetic field connectivity. The different polar fields also have impact on the resulting solar wind speed. As shown in Figure~\ref{cme_bg} and Figure~\ref{upstream}, the region of fast solar wind in the south is much wider and stronger in Case I than in Case II.

Given the many differences between the two cases resulting from the two magnetic map inputs, one could potentially use the multi-wavelength remote-sensing observations and in-situ measurements to help select the more appropriate input magnetic map, for instance, by comparing the on-disk structures shown in the EUV observations (e.g., coronal holes) or the large-scale solar wind structures (e.g., helmet streamers) seen in the white-light observations. In addition, the in-situ measurement of plasma parameters could be used to validate the solar wind solutions. However, we want to emphasize that the purpose of this study is not to distinguish which type of magnetic map is better but rather to illustrate the model uncertainty due to the imperfect magnetic field observations, which has to be taken into account whether the model is utilized for space weather prediction or for scientific research. In the meantime, we suggest that the magnetic input source should be explicitly mentioned in research papers that use global MHD models and the associated model uncertainties be discussed wherever possible. This study also emphasizes the need to have better observational coverage of the solar magnetic field for improving space weather forecasting, and should be a substantial consideration in the development of future missions. All the differences in the evolution of shock parameters could lead to significantly different inferred particle acceleration processes and result in different expected SEP spectra, complicating the interpretation of the SEP observations.

\acknowledgments
We are very grateful to the referees for invaluable comments that helped improve the paper. We thank Marc DeRosa at LMSAL for the helpful discussion on the LM synchronic magnetic maps. MJ, NVN, and CMSC are supported by NASA HSR grant 80NSSC18K1126. We thank the  The simulation results were obtained using the Space Weather Modeling Framework (SWMF), developed at the Center for Space Environment Modeling (CSEM), University of Michigan (\url{https://github.com/MSTEM-QUDA/SWMF}). We are thankful for the use of the NASA Supercomputer Pleiades at Ames and for its supporting staff for making it possible to perform the simulations presented in this paper. SDO is the first mission of NASA’s Living With a Star Program.

The LM synchronic magnetic map was downloaded directly from the \texttt{PFSS} package in \texttt{SSWIDL}. The Stanford HMI diachronic magnetic map is accessible at JSOC (\url{http://jsoc.stanford.edu/}). The AWSoM and EEGGL models used in this study are available through NASA CCMC (\url{https://ccmc.gsfc.nasa.gov/}). The spacecraft location data were downloaded from \url{https://omniweb.gsfc.nasa.gov/coho/helios/heli.html}. The simulation data used in this study is available at \url{https://10.5281/zenodo.5787007}.


%
%

\bibliography{ref}

%
%
%
%
%

\newpage
\begin{table}
\caption{Summary of Results from the Two Magnetic Maps}
\centering
\begin{tabular}{l c c}
\hline
 Magnetic Map  & LM Synchronic Map & HMI Diachronic Map\\
\hline 
 Total Unsigned Flux [10$^{23}$ Mx] & 3.5 & 4.1 \\
 Assimilating Window Unsigned Flux [10$^{23}$ Mx] & 1.3 & 1.2 \\
Flux Imbalance [\% of Total Unsigned Flux] & -2.0 & 1.3\\ 
 Mean Polar Field$^{*}$ (North/South) [Gauss] & -1.5/2.7 & -2.2/2.6  \\
 MHD Open Field Area (Positive/Negative) [10$^{11}$km$^2$] & 6.4/7.3 & 6.7/6.8\\ 
 PFSS Open Field Area (Positive/Negative) [10$^{11}$km$^2$] & 6.5/7.6 & 7.6/6.6\\
\hline
\multicolumn{3}{l}{$^{*}$The polar field regions are defined as areas with $>$60 degrees in latitude.}
\footnotetext{note}
\end{tabular}
\label{summarytable}
\end{table}

\end{document}